\documentclass[preprints,article,accept,pdftex,moreauthors]{Definitions/mdpi} 

\firstpage{1} 
\pubvolume{1}
\issuenum{1}
\articlenumber{0}
\pubyear{2025}
\copyrightyear{2025}
\datereceived{} 
\dateaccepted{ } 
\datepublished{} 
\hreflink{https://doi.org/} 
\pdfoutput=1

\usepackage{amsmath} 
\usepackage{amssymb}
\usepackage{amsfonts,dsfont} 
\usepackage{mathtools}
\usepackage{physics}
\usepackage{footmisc}

\Title{Transient Dynamics and Homogenization in Incoherent Collision Models}

\TitleCitation{Transient Dynamics and Homogenization in Incoherent Collision Models}

\Author{G\"oktu\u{g} Karpat $^{1,}$*\orcidA{} and Bar\i\c{s} \c{C}akmak$^{2,}$*\orcidB{}}

\AuthorNames{G\"oktu\u{g} Karpat and Bar\i\c{s} \c{C}akmak}

\AuthorCitation{Karpat, G.; \c{C}akmak, B.}
 
\address{$^{1}$ \quad Faculty of Engineering and Natural Sciences, Sabanci University, Tuzla, Istanbul 34956, Türkiye \\

$^{2}$ \quad Department of Physics, Farmingdale State College—SUNY, Farmingdale, NY 11735, USA}

\corres{Correspondence: goktug.karpat@sabanciuniv.edu (GK); cakmakb@farmingdale.edu (B\c{C})}

\abstract{Collision models have attracted significant attention in recent years due to their versatility to simulate open quantum systems in different dynamical regimes. They have been used to study various interesting phenomena such as the dynamical emergence of non-Markovian memory effects and the spontaneous establishment of synchronization in open quantum systems. In such models, the repeated pairwise interactions between the system and the environment and also the possible coupling between different environmental units are typically modeled using the coherent partial-swap (PSWAP) operation as it is known to be a universal homogenizer. In this study, we investigate the dynamical behavior of incoherent collision models, where the interactions between different units are modeled by the incoherent controlled-swap (CSWAP) operation, which is also a universal homogenizer. Even though the asymptotic dynamics of the open system in case of both coherent and incoherent swap interactions appear to be identical, its transient dynamics turns out to be significantly different. Here we present a comparative analysis of the consequences of having coherent or incoherent couplings in collision models, namely, PSWAP or CSWAP interactions respectively, for the emergence of memory effects for a single qubit system and for the onset synchronization between a pair of qubits, both of which are strictly determined by the transient dynamics of the open system. 
}


\begin{document}
\section{Introduction}

Collision models~\cite{Ciccarello2017,Ciccarello2021,Campbell2021,Cusumano2022}, which are now an indispensable tool to simulate open quantum systems, were first introduced nearly $60$ years ago to study thermalization of a quantum system through sequential interactions with reservoir particles at a certain temperature~\cite{Rau}. Generalization of this idea to study the interaction between a system and reservoir particles in arbitrary states, and analyze the spread of system information throughout the reservoir came in the early years of the $21^{st}$ century~\cite{Ziman2002,Scarani2002,Nagaj2002,Ziman2005,Ziman2010,Cusumano2018}. These works named the process as \emph{quantum homogenization}, since the reservoir particles are no longer represented by thermal states, and sequential interaction of the system particle with identically prepared non-equilibrium reservoir particles will eventually \emph{equilibriate} the system state with the reservoir state. In Ref.~\cite{Ziman2002}, partial SWAP (PSWAP) operation was shown to be the unique and universal homogenizer among all possible unitaries acting on a single system and a reservoir, such that the state of the system can be brought arbitrarily close to and have negligible effect on the state of the reservoir after many sequential applications, measured by an appropriate distance measure between quantum states. It is important to note that PSWAP leads to a coherent interaction between the system and the reservoir. Recently, an incoherent universal quantum homogenizer was introduced in Ref.~\cite{Beever2024}, based on a controlled SWAP (CSWAP) gate between system and reservoir qubits that achieves homogenization with a very similar convergence rate as the PSWAP model. 

Aforementioned initial studies paved the way to the heavy utilization of collision (or repeated interaction) models across very different fields of research~\cite{lorenzo2015a,lorenzo2015b,Strasberg2017,pezzutto2016,Cusumano2017,Cusumano2018-2,Cilluffo2020,pezzutto2019,abah2019,beyer2018,baldijao2018,campbell2019a,Rodrigues2019,Karpat2019,Guarnieri2020,Cattaneo2021}. The reason behind such increasing popularity is, we believe, two-fold. First, in contrast to traditional techniques that are used to deal with open quantum systems~\cite{BreuerPetruccione,AlickiLendiBook,deVega2017}, collision models offer a highly accessible framework both conceptually and theoretically. The accessibility of the method stems from its construction, in which the environment is represented as a collection of particles that sequentially interact with the system particle. In addition, the flexibility in determining the nature (Hamiltonian), order, and timing of interactions offers a versatile framework for examining complex dynamical quantum processes. One of the most technically challenging tasks in the theory of open quantum systems is to deal with problems where it is not possible to resort to the Markovian approximation~\cite{deVega2017}, i.e. non-Markovian processes. The ways of characterizing and quantifying memory effects from a quantum information perspective have been a topic of intense debate for more than a decade now~\cite{Rivas2014,nonmarkovRMP}. Owing to their versatility, collision models provided a natural test-bed for memory effects in open quantum systems~\cite{Ciccarello2013,mccloskey2014,vacchini2014,Bernandes2014,Kretschmer2016,Cakmak2017,lorenzo2017a,lorenzo2017b,filippov2017,campbell2018,jin2018,campbell2019b,cakmak2021,Karpat2021}. Recently, it has also been shown that non-Markovian collision models, whose non-Markovianity is solely based on the interaction between reservoir particles that transmit system information to particles that are yet to interact with the system particle, are also capable of homogenizing the quantum system to the state of identically prepared reservoir units~\cite{Saha2024}. However, in the case of locally identical but globally correlated reservoir states, which also bring in memory due to the presence of correlations, homogenization is hindered~\cite{Comar2021}, and the structure of the generated multipartite correlations among the constituents of the collision model are highly non-trivial~\cite{Filippov2021,Cusumano2024,Purkayastha2021,Purkayastha2022}. To the best of our knowledge, homogenization of a bipartite or multipartite system, has not been investigated in the literature.

The aim of this work is to analyze the differences between the transient dynamics generated by the well-established coherent PSWAP~\cite{Ziman2002} and the recently introduced incoherent CSWAP~\cite{Beever2024} universal quantum homogenizers in a collision model setting with minimal ingredients. To that end, we focus on the dynamically induced memory effects and synchronization for single and two qubit systems, respectively. The particular reason why we consider these two phenomena is because both non-Markovianity and synchronization between the observables of quantum systems, develop during the transient time, before the system reaches the fixed point of the open system dynamics. Even though we observe that both PSWAP and CSWAP based models do homogenize the system particle(s), i.e. they end up equilibrating with the reservoir state, we demonstrate that the dynamical processes generated by these two approaches have, in fact, stark differences.

The paper is organized as follows. In Section~\ref{models}, we introduce the basics of our PSWAP and CSWAP based collision models together with different figures of merit that immediately point out that these models take the state of the system through very different paths on the Bloch sphere as they homogenize it. We then turn our attention to non-Markovianity in Section~\ref{nonmarkov}, and the layout which types of dynamics induce memory effects. In Section~\ref{sync}, we enlarge our system to two qubits interacting with the same bath, and try to assess the conditions under which synchronization between the qubits emerge in transient time before they homogenize with the reservoir. We conclude in Section~\ref{conc}.

\section{Coherent and Incoherent Collision Models}\label{models}

In this section, our aim is to present the fundamentals of different types of collision models, namely, coherent and incoherent collision models, that we consider in our analysis. In the collision model framework, time-evolution of an open quantum system is simulated through brief sequential interactions of the system $s$ with an environment consisting of a stream of quantum systems $e_i$, all in an identical state. It is well-known that the collision model approach provides a great deal of versatility to model open system dynamics since it allows the simulation of different dynamical regimes, such as Markovian and non-Markovian evolutions. In our treatment, we assume that both the principal open system of interest and the environmental systems are two-level quantum systems, i.e. qubits. It is also supposed that there is no previously established correlation of either quantum or classical nature between the open system qubit and the environmental qubits.

In general, in the collision models considered in the literature, the pairwise interactions between the open system qubit $s$ and the environment qubits $e_i$, and also the possible couplings between different environmental units $e_i$ and $e_{i+1}$, take place coherently in the sense that the resulting time-evolution operator is unitary, that is, it can be generated by a Hamiltonian model. Typically, this unitary operator is taken to be the PSWAP operation,
\begin{equation}
U_{c}(\gamma)=\cos\gamma \mathbb{I}_4 + i \sin\gamma \mathbb{S},
\end{equation}
where $\mathbb{I}_4$ is the $4\times4$ identity matrix and $\mathbb{S}$ denotes the SWAP operator,
\begin{equation}
\mathbb{S}=|00\rangle\langle00|+|01\rangle\langle10|+|10\rangle\langle01|+|11\rangle\langle11|,
\end{equation}
written in the computational basis with $\gamma$ being a dimensionless real number quantifying the strength of the interaction. We would like to note that PSWAP unitary can be generated under the evolution of energy preserving interactions, such as the Heisenberg and effective dipole-dipole Hamiltonians, with $\gamma$ being directly proportional to the interaction strength between the systems to be swapped~\cite{Ziman2002,pezzutto2016,cakmak2019}. In this work, we denote the strength of the pairwise PSWAP couplings between the qubit $s$ and the environment qubits $e_i$, and the intra-environmental interactions are respectively characterized by $\gamma_{se}$ and $\gamma_{ee}$. The main reason for the choice of PSWAP interaction operator is that, in the absence of intra-environmental interactions, such a model with weak system-environment coupling fulfills the requirements for a universal quantum homogenizer, i.e., any system qubit state will be transformed to any chosen fixed environmental qubit state, leaving the state of the environment approximately unchanged~\cite{Ziman2002}. Note that the generalization of this protocol to $d$-level quantum systems has also been introduced and an all-optical test of the protocol is proposed for Gaussian states~\cite{Nagaj2002}.

Let us now describe the details of a single step in the coherent collisional dynamics. In the first step, the system qubit $s$ interacts with the first environment qubit $e_1$ through a PSWAP interaction described by $U_{c}(\gamma_{se})$. Afterwards, the environment particle $e_1$, which interacted with the system qubit $s$ previously, interacts with the next environment qubit $e_2$ via a PSWAP operation $U_{c}(\gamma_{ee})$. Lastly, a single step in the evolution is completed by tracing out the environment qubit $e_1$, and then the same procedure is iteratively repeated, restarting with the coupling of the system qubit $s$ and the environment qubit $e_2$. When the interactions between the environmental units are assumed to vanish ($\gamma_{ee}=0$), dynamics of the open system becomes memoryless and thus Markovian, since there is no mechanism for the system to recover the information lost to the environment during the time evolution.

From a different point of view, one can consider the interaction between the system qubit $s$ and the environment qubits $e_i$, and also the potential couplings between the environmental qubits $e_i$ and $e_{i+1}$ to be incoherent. In particular, here we will focus on the CSWAP operation, which has been very recently used to model a universal incoherent homogenizer, as an alternative to the coherent PSWAP homogenizer~\cite{Ziman2002,mccloskey2014}. As opposed to the two-qubit PSWAP operation, the CSWAP is a three-qubit operation, where the interaction between the second and third qubits is mediated by the first qubit, which serves as a control system. Specifically, the CSWAP operation can be written as 
\begin{equation}
U_{ic}=\frac{1}{2}(|0\rangle\langle0| \otimes \mathbb{I}_4 + |1\rangle\langle1| \otimes \mathbb{S}).
\end{equation}
This clearly implies that if the first (control) qubit is in the state $|0\rangle$, the second and the third qubits evolve trivially, but if the control qubit is in the state $|1\rangle$, the the two-qubit SWAP operation acts on them. In our study, we choose the state of the control qubit $c$ as
\begin{equation}
|c\rangle=\cos\gamma |0\rangle + \sin\gamma |1\rangle,
\end{equation}
which is a weighted superposition of the states $|0\rangle$ and $|1\rangle$. Here, the parameter $\gamma$ determines the strength of the SWAP interaction between the second and the third qubits.

Although the action of PSWAP and CSWAP interaction operators on a two-qubit system, e.g., on the system qubit $s$ and the first environment qubit $e_1$ are quite similar, there is one crucial difference in the post-interaction state of the system qubit. Particularly, as shown in Ref.~\cite{Ziman2002,Beever2024}, after a single PSWAP interaction between the system qubit $s$ and the environment qubit $e_1$, the state of the system qubit is described by 
\begin{equation}\label{s-pswap}
\rho_s=\frac{1}{2}\left[\mathbb{I}_2 + \cos^2\gamma (\Vec{\beta} \cdot \Vec{\sigma}) + \sin^2\gamma (\Vec{\alpha} \cdot \Vec{\sigma}) + \frac{\cos\gamma \sin\gamma }{4} (\Vec{\beta} \cross \Vec{\alpha}) \cdot \Vec{\sigma}  \right],
\end{equation}
where $\mathbb{I}_2$ is the $2 \times 2$ identity matrix, and $\Vec{\beta}$ and $\Vec{\alpha}$ stand for the Bloch vectors for the initial state of the system qubit $s$ and the environment qubit $e_1$ before the interaction, respectively, and $\Vec{\sigma}$ is the vector of standard Pauli operators, $\sigma_x$, $\sigma_y$, and $\sigma_z$. On the other hand, if a single CSWAP interaction takes place between the qubits $s$ and $e$, mediated by the control qubit $c$, the state of the system qubit after the interaction takes the form
\begin{equation}\label{s-cswap}
\rho_s=\frac{1}{2}\left[\mathbb{I}_2 + \cos^2\gamma (\Vec{\beta} \cdot \Vec{\sigma}) + \sin^2\gamma (\Vec{\alpha} \cdot \Vec{\sigma})  \right],
\end{equation}
A direct  comparison between Equations (\ref{s-pswap}) and (\ref{s-cswap}) reveals that the term, involving the contribution originated from the interference between the qubits $s$ and $e$, is lacking in case of the CSWAP operation, which reflects the incoherent nature of this interaction.

In the collision model scheme that we described above, one can model the system-environment and intra-environment interactions, respectively characterized by the coupling parameters $\gamma_{se}$ and $\gamma_{ee}$, using coherent ($U_c$) or incoherent ($U_{ic}$) interaction operators. Naturally, there are four different ways that the model can be constructed. In the first case, both system-environment and intra-environment interactions are described by PSWAP, where all the interactions in the model are coherent. In the second case, both interactions are described by CSWAP, which produces a collision model with fully incoherent couplings. In the third (fourth) case, as the system-environment interaction is generated by PSWAP (CSWAP), intra-environment coupling is represented by CSWAP (PSWAP), which gives rise to a collision model, involving a mixture of coherent and incoherent interactions. 

For concreteness, let us explicitly write down the discrete evolution of the open system $s$ in case both system-environment and intra-environment couplings are described by the coherent PSWAP interactions. The density operator of the open system qubit $s$ after the $i$th step of the collision model  $\rho_s[i+1]$ can be calculated as
\begin{equation}
\rho_s[i+1]=\Tr_{e_i e_{i+1}}\left\{ U_c(\gamma_{ee}) U_c(\gamma_{se})  \left( \rho_s[i] \otimes \Tilde{\rho}_{e_{i}}[i] \otimes\rho_{e_{i+1}}[i] \right)  U^\dagger_c(\gamma_{se}) U^\dagger_c(\gamma_{ee})    \right\},
\end{equation}
where $\Tr_{e_i e_{i+1}}$ denotes the partial trace operation over the environmental qubits $e_i$ and $e_{i+1}$. Also, $\Tilde{\rho}_{e_{i}}[i]=\Tr_s \left\{  \rho_{se_i}[i]\right\}$, which is the partial trace over the system qubit $s$, still contains some information about the system qubit as a result of its interaction with the $(i+1)$th environment in the previous iteration. The state of the open system in the remaining three cases of different PSWAP and CSWAP combinations can be obtained similarly, replacing $U_c$ by $U_{ic}$ and introducing a control qubit $\rho_c=|c\rangle \langle c|$ for incoherent couplings.

We will commence our analysis by focusing on the case, where there is no coupling between environmental qubits ($\gamma_{ee}=0$), i.e., memoryless evolutions, and comparatively discuss the consequences of having coherent (PSWAP) and incoherent (CSWAP) interactions between the system qubit $s$ and the environmental qubits $e_i$. In all the collision models in our work, we set the initial state of the identical environmental qubits to be $\rho_{e_i}=|0\rangle \langle 0|$. Besides, for single qubit models, unless otherwise stated, the open system $s$ is assumed to start out in an equal superposition of $|0\rangle$ and $|1\rangle$, that is, in the state $|+\rangle= (1/\sqrt{2})(|0\rangle+|1\rangle)$, and both for PSWAP and CSWAP couplings between the system qubit $s$ and the environment qubits $e_i$, we suppose that the strength of the coupling is $\gamma_{se}=0.05 (\pi/2)$.

Fidelity between two density operators $\rho_1$ and $\rho_2$ can be employed to quantify the degree of similarity between two quantum systems, whose states are represented by these density operators. While the fidelity of orthogonal states vanishes, fidelity acquires its maximum value of one for identical states. For a pair of qubits, it is defined as~\cite{nielsen}
\begin{equation}\label{fid}
F(\rho_1, \rho_2)= \Tr{\rho_1\rho_2} + 2\sqrt{\det\rho_1\det\rho_2},
\end{equation}
where $\Tr$ and $\det$ respectively denote the trace and the determinant operations. Furthermore, von-Neumann entropy of the density operator $\rho$ reads
\begin{equation}\label{entrpy}
S(\rho)=- \Tr{\rho \ln \rho}.
\end{equation}
Lastly, the amount of quantum coherence contained in a qubit state, represented by the density operator $\rho$, can be calculated using the widely-used $l_1$ norm of coherence~\cite{Baumgratz2014,Streltsov2017},
\begin{equation}\label{chrnc}
C(\rho)=\sum_{k\neq l}|\langle k |\rho| l \rangle|,
\end{equation}
which is simply the sum of the absolute values of
the off-diagonal elements of $\rho$.

\begin{figure}[t]
\centering
\includegraphics[width=0.85\columnwidth]{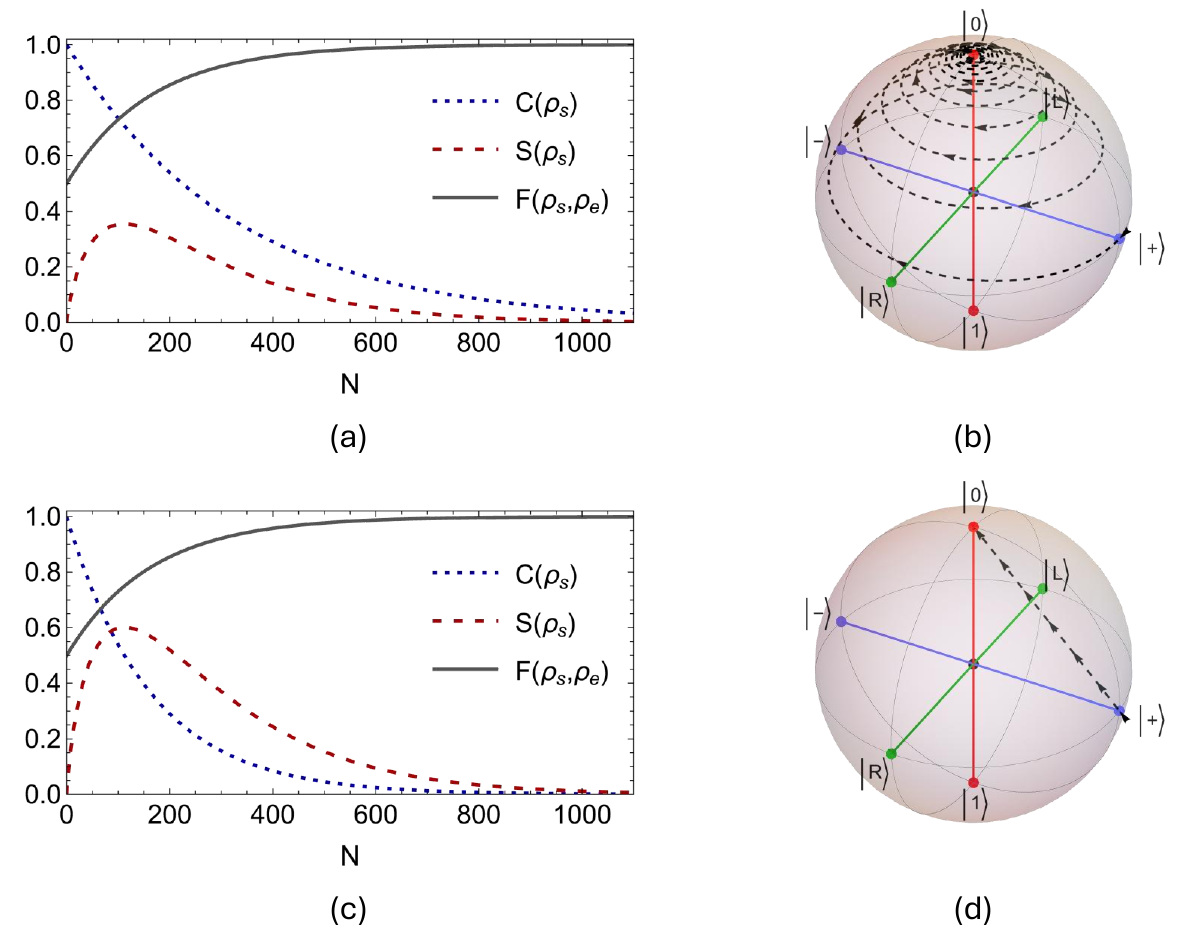}
\caption{For the coherent PSWAP interaction between the system qubit $s$ and the environment qubits $e_i$ without intra-environment couplings, $\gamma_{ee}=0$ and $\gamma_{se}=0.05 (\pi/2)$, (a) shows the evolution of the coherence $C(\rho_s)$ in the open system, the entropy $S(\rho_s)$ of the open system, and the fidelity $F(\rho_s, \rho_e)$ between the open system and the initial state of the environment qubits $\rho_e=|0\rangle \langle 0|$ for $N=1100$ collisions. (b) displays the path of the open system state $s$ through the Bloch ball starting from the state $\rho_s=|+\rangle \langle +|$. On the other hand, (c) and (d) display the same set of plots as in (a) and (b) when the interaction between the system and the environment is described by the incoherent CSWAP coupling with the same interaction parameters, that is, $\gamma_{ee}=0$ and $\gamma_{se}=0.05 (\pi/2).$}
\label{fig1}
\end{figure}

In Figure \ref{fig1}, we examine the dynamics of quantum coherence $C(\rho_s)$ and von-Neumann entropy $S(\rho_s)$ of the system qubit, along with the evolution of the fidelity $F(\rho_s, \rho_e)$ between the system qubit state $\rho_s$ and the state of the environment qubits $\rho_e=|0\rangle \langle 0|$ for $N=1100$ collisions, in cases of Markovian (a) coherent and (c) incoherent collision models, where intra-environment interactions vanish. At the same time, Figures \ref{fig1} (b) and (d) display the evolution path for the Bloch vector of the system qubit through the Bloch ball for coherent and incoherent collision models, respectively. Looking at the dynamics of the fidelity $F(\rho_s, \rho_e)$ between the evolving system state and the environment state $\rho_e=|0\rangle \langle0|$ in Figures \ref{fig1} (a) and (c), it is straightforward to see that both coherent and incoherent Markovian collision models converge to their identical final state $|0\rangle$ after almost exactly the same number of collisions. Nonetheless, the transient dynamics of the initial system state $\rho_s=|+\rangle \langle+|$ under these models is significantly different, as clearly demonstrated by the evolution of Bloch vectors. Whereas the Bloch vector in case of the incoherent model, which is realized by CSWAP couplings, follows a straight line on the $xz$-plane toward the state $|0\rangle$, it instead spirals through the Bloch ball towards the same state when it comes to the coherent model involving PSWAP interactions. We stress that the observed behavior here is not specific to the chosen initial system state $|+\rangle$, but rather a general feature of the coherent and incoherent collision models considered in our work. Besides, we observe that when the system-environment coupling is incoherent, the system qubit $s$ loses its coherence $C(\rho_s)$ at a slightly faster rate and its entropy $S(\rho_s)$ reaches a higher maximum value in early dynamics, as compared to the incoherent case. These two distinct dynamical behaviors can be understood by noticing the differences in the evolution path of system states in the Bloch ball as demonstrated in Figures \ref{fig1} (b) and (d). Lastly, considering that both coherent and incoherent collision models converge to their final state after almost exactly the same number of collisions, it is also interesting to notice that the time-evolution of the system $s$ in some sense occurs at a much faster rate for PSWAP interaction in comparison to the CSWAP interaction, since the path taken by the Bloch vector is much longer for the former. We would like to note that all Bloch sphere plots are generated using the \verb|melt| library~\cite{melt}.

\section{Quantum non-Markovianity}\label{nonmarkov}

The present section is devoted to the comparative examination of memory effects, which are dynamically emerging during the evolution of the open system due to the non-Markovian behavior, for coherent and incoherent collision models. In particular, for our anaylsis in this section, we consider four different collision model settings. Namely, if both the system-environment coupling and the intra-environmental interactions are coherent (incoherent), we will refer to the model as PSWAP-PSWAP (CSWAP-CSWAP) collision model. In addition, when the system-environment coupling is coherent (incoherent) and the intra-environmental interactions are incoherent (coherent), we will refer to the model as PSWAP-CSWAP (CSWAP-PSWAP) collision model.

Let us first discuss how we identify non-Markovianity in the dynamics of open quantum systems and quantify the degree of memory effects in non-Markovian evolutions. Suppose that a quantum process $\Lambda(t,0)$, that is, a completely positive trace-preserving (CPTP) map, represents the dynamical evolution of an open quantum system. Conventionally, a quantum process is said to be Markovian, or equivalently memoryless, provided that it satisfies the property of CP-divisibility, which implies that decomposition rule $\Lambda(t,0)=\Lambda(t,s)\Lambda(s,0)$ holds with $\Lambda(t,s)$ being a CPTP map for all $s\leq t$. Indeed, there exist legitimate quantum processes that violate the decomposition rule, in which case $\Lambda(t,s)$ fails to be a CPTP map or even fails to exist. In such situations, the process $\Lambda(t,0)$ is called to be CP-indivisible and thus non-Markovian, exhibiting memory effects.

\begin{figure}[t]
\centering
\includegraphics[width=0.85\columnwidth]{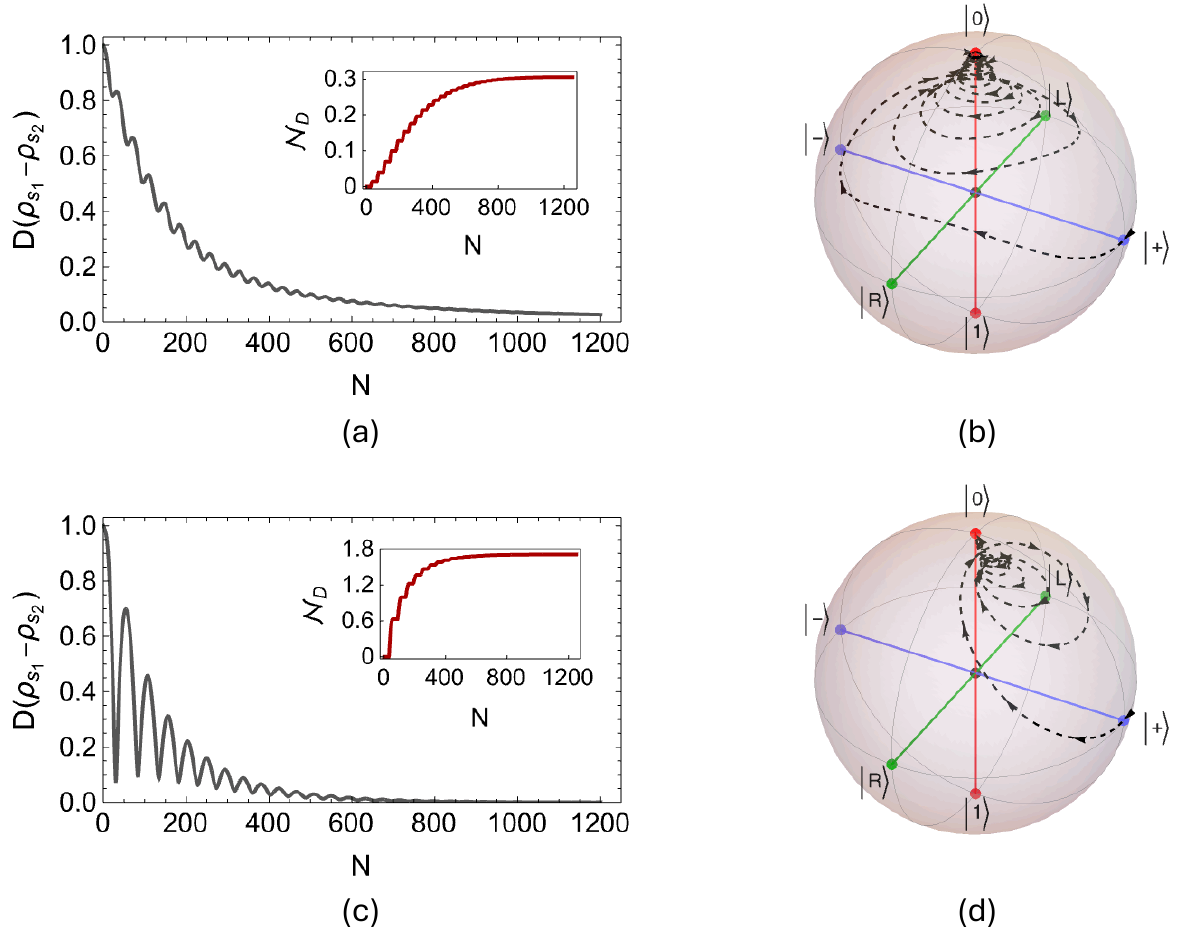}
\caption{While (a) shows the dynamics of the trace distance for the PSWAP-PSWAP collision model, together with the evolution of the non-Markovianity measure $\mathcal{N}_D$ shown in the inset for $N=1200$ collisions, (c) displays the same set of plots in case of the PSWAP-CSWAP collision model for same number of collisions. We take the initial system state pair as $|\pm\rangle=(1/\sqrt{2})(|0\rangle\pm|1\rangle)$, and set the system-environment and the intra-environment coupling strengths identically in both models as $\gamma_{se}=0.05 (\pi/2)$ and $\gamma_{ee}=0.93 (\pi/2)$. In (b) and (d), we demonstrate the paths of evolution of the Bloch vectors through the Bloch ball for the PSWAP-PSWAP and PSWAP-CSWAP models, respectively, starting from the initial system state $\rho_s=|+\rangle \langle +|$, with the same interaction parameters.}
\label{fig2}
\end{figure}

In order to measure the degree of memory in the time-evolution of the open system, we consider the most widely used quantifier of non-Markovianity in the literature, i.e., the Breuer-Laine-Piilo (BLP) measure, which is based on the distinguishability of open system states throughout the dynamics of the system~\cite{BreuerMeasureDegree2009}. In particular, BLP measure proposes that signatures of non-Markovian memory effects can be captured via the dynamical behavior of the trace distance that quantifies to what extent two open system states can be distinguished during the time-evolution. Trace distance between the density operators $\rho_1$ and $\rho_2$ reads
\begin{equation}
D(\rho_1, \rho_2)= \frac{1}{2} ||\rho_1-\rho_2||_1= \frac{1}{2} \Tr [(\rho_1-\rho_2)^{\dagger} (\rho_1-\rho_2)]^{1/2},
\end{equation}
where $||.||_1$ denotes the trace norm. As $D(\rho_1, \rho_2)$ reaches its maximum value of one when $\rho_1$ and $\rho_2$ are orthogonal, it vanishes for an identical pair of density operators. In the BLP approach, the dynamics of the trace distance and thus of distinguishability between two arbitrary initial open system states is interpreted as the flow of information between the open system and its environment. Specifically, when $dD/dt<0$ throughout the dynamics, indicating that distinguishability between two arbitrary open system states decreases at all times, the evolution is said to be Markovian since in this case information flows from the system to the environment monotonically. Conversely, a temporary increase in the trace distance, $dD/dt>0$ during some time interval, signifies a flow of information from the environment back to the open system, which in turn gives rise to non-Markovian behavior. As a consequence, following the BLP approach, the degree of non-Markovianity of an open quantum system evolution is measured based on the quantity~\cite{BreuerMeasureDegree2009}
\begin{equation}
\mathcal{N}_D= \max_{\rho_1(0),\rho_2(0)} \int_{\dot{D}>0}\frac{dD}{dt}dt,
\end{equation}
where the maximization is performed over antipodal pairs of initial states $\rho_1(0)$ and $\rho_2(0)$ of the open system on Bloch sphere~\cite{optimalpair}.  At this point, we emphasize that the monotonic decay of trace distance is not equivalent to CP-divisibility. That is, despite $\mathcal{N}_D=0$ for all CP-divisible dynamics, there are CP-indivisible evolutions under which trace distance does not exhibit any revival and thus $\mathcal{N}_D$ vanishes all the same. However, in such cases, the trace distance based quantity $\mathcal{N}_D$ can be considered as a measure of memory effects in the open system dynamics on its own. Due to the fact that we study collision models to simulate open system dynamics, we consider a discretized version of $\mathcal{N}_D$ as in Ref.~\cite{Laine2010},
\begin{equation} \label{tdmeasure}
\mathcal{N}_D= \max_{\rho_{1}[0],\rho_{2}[0]} \sum_i \left[D(\rho_{1}[i],\rho_{2}[i])-D(\rho_{1}[i-1],\rho_{2}[i-1]) \right].
\end{equation}

We begin our investigation studying dynamical memory effects in the two collision models with coherent system-environment interactions. To put it differently, we will first consider the PSWAP-PSWAP and PSWAP-CSWAP collision models with weak system-environment coupling, where intra-environment interactions are incoherent in the latter. We numerically confirm that in both cases the optimal open system state pair that achieves the maximum in Equation (\ref{tdmeasure}) is given by the orthogonal operators $\rho_1[0]=|+\rangle\langle+|$ and $\rho_2[0]=|-\rangle\langle-|$, where $|\pm\rangle=(1/\sqrt{2})(|0\rangle\pm|1\rangle)$. In addition, we fix the strength of the system-environment and the intra-environment interactions as $\gamma_{se}=0.05 (\pi/2)$ and $\gamma_{ee}=0.93 (\pi/2)$, respectively, in both of the considered models. In Figures \ref{fig2} (a) and (c), we display the dynamics of the trace distance for $N=1200$ collisions respectively for the PSWAP-PSWAP and PSWAP-CSWAP models, along with the degree of non-Markovianity $\mathcal{N}_D$, which are shown as insets. It is clear that, when the pairwise couplings between the system $s$ and the environment qubits $e_i$ are realized by the coherent PSWAP operator, incoherent CSWAP interactions between the environment qubits $e_i$ and $e_{i+1}$ give rise to a higher degree of memory in the dynamics, as compared to having coherent PSWAP intra-environment interactions. Moreover, in Figures \ref{fig2} (b) and (d), we show the evolution paths of the Bloch vectors starting from the initial state $\rho_s=|+\rangle \langle +|$ for the PSWAP-PSWAP and PSWAP-CSWAP models, respectively. We observe that in both cases the initial state spirals up to the same final state $|0\rangle$, however, the routes that the Bloch vectors follow are distinct. In fact, in comparison to the Bloch vector evolution of the Markovian PSWAP model shown in Figure \ref{fig1} (b), the paths in Figures \ref{fig2} (b) and (d) have an important difference. That is, as the $z$ component of the Bloch vector increases monotonically for the Markovian PSWAP model, it evolves toward the north pole in a non-monotonic fashion for the non-Markovian PSWAP-PSWAP and PSWAP-CSWAP models.

\begin{figure}[t]
\centering
\includegraphics[width=0.85\columnwidth]{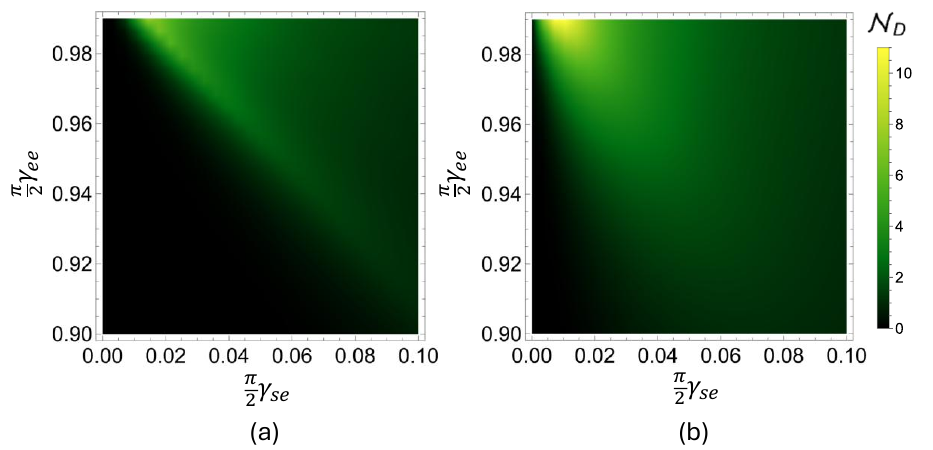}
\caption{Non-Markovianity diagrams for (a) the PSWAP-PSWAP and (b) the PSWAP-CSWAP collision models in terms of the system-environment and the intra-environment interaction parameters, $\gamma_{se}$ and $\gamma_{ee}$. For both models, we simulate the dynamics for $N=12000$ iterations and the state pair used in the calculation of the non-Markovianity measure $\mathcal{N}_D$ is fixed as $|\pm\rangle=(1/\sqrt{2})(|0\rangle\pm|1\rangle)$. }
\label{fig3}
\end{figure}

Having seen that incoherent interactions between the environment qubits $e_i$ and $e_{i+1}$ can enhance amount of information backflow from the environment to the open system, resulting in a higher degree of memory, we examine the behavior of the non-Markovianity measure $\mathcal{N}_D$ for a wider range of coupling strengths. In Figure \ref{fig3}, we present the degree of non-Markovianity of the system qubit $s$, calculated for $N=12000$ collisions, after which point the qubit $s$ reaches its final state for the considered system-environment and intra-environment interaction parameters range, i.e., $\gamma_{se}=0.00-0.10$ and $\gamma_{se}=0.90-0.98$. Initial state pair for the open system qubit is once again set to $|\pm\rangle=(1/\sqrt{2})(|0\rangle\pm|1\rangle)$. While Figure \ref{fig3} (a) displays the non-Markovianity measure $\mathcal{N}_D$ for the PSWAP-PSWAP model, in terms of the strength of system-environment and intra-environment interactions, namely $\gamma_{se}$ and $\gamma_{ee}$, Figure \ref{fig3} (b) shows the results of the same analysis in case of the PSWAP-CSWAP model. We see that, in general, incoherent intra-environment couplings realized by the CSWAP interaction enhances the degree of memory effects as compared to the case of coherent intra-environment couplings generated by the PSWAP operator. Indeed, for a given non-zero system-environment coupling strength $\gamma_{se}$, unless the pairwise interactions between the environment qubits $e_i$ and $e_{i+1}$ are sufficiently strong, the PSWAP-PSWAP model does not exhibit memory effects. To put it differently, there is a considerable range of interactions parameters for which the PSWAP-PSWAP model remains in the Markovian regime, while the PSWAP-CSWAP model becomes non-Markovian.

\begin{figure}[t]
\centering
\includegraphics[width=0.85\columnwidth]{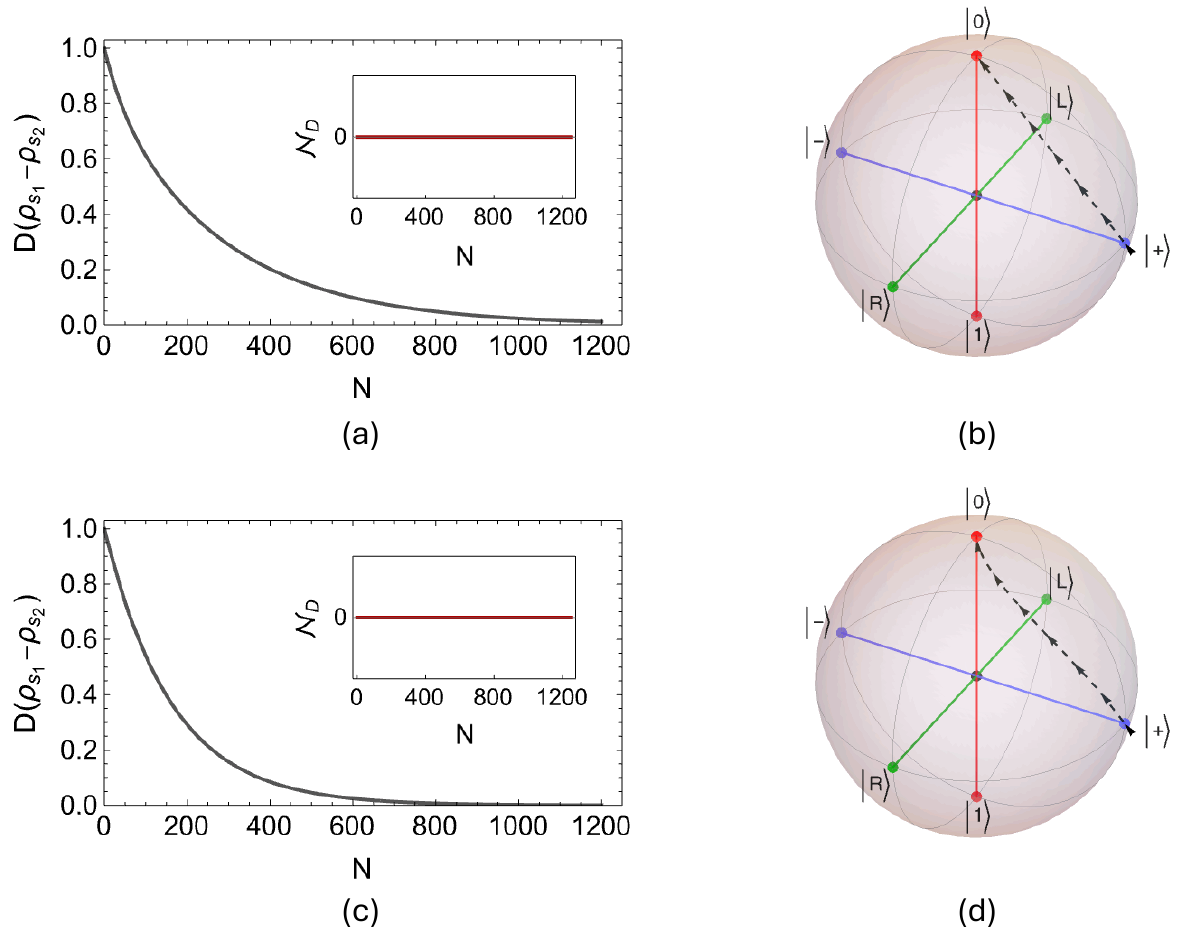}
\caption{While (a) shows the dynamics of the trace distance for the CSWAP-CSWAP collision model, together with the evolution of the non-Markovianity measure $\mathcal{N}_D$ shown in the inset for $N=1200$ collisions, (c) displays the same set of plots in case of the CSWAP-PSWAP collision model for same number of collisions. We take the initial system state pair as $|\pm\rangle=(1/\sqrt{2})(|0\rangle\pm|1\rangle)$, and set the system-environment and the intra-environment coupling strengths identically in both models as $\gamma_{se}=0.05 (\pi/2)$ and $\gamma_{ee}=0.93 (\pi/2)$. In (b) and (d), we demonstrate the paths of evolution of the Bloch vectors through the Bloch ball for the CSWAP-CSWAP and CSWAP-PSWAP models, respectively, starting from the system state $\rho_s=|+\rangle \langle +|$, with the same interaction parameters.}
\label{fig4}
\end{figure}

Lastly, we turn our attention to the remaining two collision models, namely CSWAP-CSWAP and CSWAP-PSWAP models, where in both cases the couplings between the system qubit $s$ and environment qubits $e_i$ are incoherent. Considering the same model parameters as in the case of two previously analyzed collision models, that is, $\gamma_{se}=0.05 (\pi/2)$ and $\gamma_{ee}=0.93 (\pi/2)$, we perform a numerical search over all antipodal pairs of system states on the Bloch sphere. We find that it is not possible to witness temporary increases in the dynamics of the trace distance for any of the numerically traced state pairs. Hence, we conclude that the evolution realized by both CSWAP-CSWAP and CSWAP-PSWAP collision models is Markovian. In Figure \ref{fig4} (a) and (c), we show the time-evolution of trace distance for the pair $|\pm\rangle=(1/\sqrt{2})(|0\rangle\pm|1\rangle)$ respectively for CSWAP-CSWAP and CSWAP-PSWAP models to exemplify the monotonic decay of trace distance and thus memoryless evolution in both cases. In fact, we have also extended our numerical search to study the limit, where the intra-environment interaction strength approaches to its maximum value, $\gamma_{ee} \rightarrow \pi/2$, to look for a possible backflow of information from the environment to the open system. However, although increasing values of $\gamma_{ee}$ cause the system qubit to reach its final state after more and more pairwise collisions, this does not change the monotonic behavior of trace distance. Also, Figure \ref{fig4} (b) and (d) demonstrate the evolution of the Bloch vector path for the system state, where $z$ components increase monotonically for both models, as in case of memoryless models with no intra-environment collisions.

Let us briefly summarize our findings in this section. Under weak system-environment coupling assumption, our simulations establish that the emergence of memory effects in open system dynamics is intimately related to having a coherent coupling through the PSWAP operator between the system and its environment. Incoherent system-environment interactions, which are realized by the CSWAP operator, always leads to Markovian and thus memoryless evolutions, independently of the coherence or incoherence of the pairwise interactions between the environmental qubits. Nevertheless, in case the system-environment interaction takes place coherently via the PSWAP operator, having incoherent pairwise CSWAP couplings rather than coherent PSWAP interactions between the environment qubits generally amplifies the degree of non-Markovianity of the dynamics.

Finally, we would like to discuss a technical issue on the universality of the homogenization process in the considered non-Markovian collision models. It has very recently been shown that PSWAP-PSWAP model is a universal homogenizer~\cite{Saha2024}, although one can obtain higher degrees of non-Markovianity by considering different intra-environment interactions. Nevertheless, there is no direct proof that a non-Markovian model that includes at least one CSWAP interaction in its construction is a universal homogenizer. However, the difference between the system-reservoir fidelities between Markovian PSWAP and CSWAP homogenizers at any time during the dynamics is at most $\%2$ and gets smaller further into the dynamics~\cite{Beever2024}. This implies that the convergence rate of the system state towards the environment state is nearly identical in the large reservoir limit. While this result do not necessarily imply anything with certainty in the non-Markovian limit, evaluating Refs.~\cite{Saha2024} and~\cite{Beever2024} together points towards a direction, where non-Markovian models built with a combination of PSWAP and CSWAP interactions may also be universal.

\section{Quantum synchronization}\label{sync}

In this section, we examine the consequences of having coherent versus incoherent interactions for the establishment of environment-induced spontaneous synchronization~\cite{Giorgi2013,Giorgi2016,Bellomo2017,Cabot2019,Karpat2019,Karpat2020,Cabot2021,Galve2017,Giorgi2019} between a pair of qubits using the collision model framework. As the discussion of quantum synchronization requires us to consider two system particles, we first introduce the collision model setting to be used to simulate the open system evolution of a two-qubit system. As in the case of the single qubit open system model, which has been previously used to study the behavior of non-Markovianity, here we also consider both coherent PSWAP and incoherent CSWAP interactions between the system particle and the environment units in the collision model framework.
We now give the details of a single iteration step in the fully coherent collision model that we intend to analyze in this part, assuming a PSWAP interaction between both systems and environment, i.e. a PSWAP-PSWAP model. The collision scheme starts with the interaction of the first system qubit $s_1$ with the first environment qubit $e_1$ through a PSWAP interaction $U_{c}(\gamma_{s_1e})$. Then, the second qubit $s_2$ interacts with the same environment qubit $e_1$ with an identical interaction operator $U_{c}(\gamma_{s_2e})$. Following these, the system qubits $s_1$ and $s_2$ evolve freely with their self-evolution operators $U_{s_1}=\exp{-iH_{s_1}\delta t_s}$ and $U_{s_2}=\exp{-iH_{s_2}\delta t_s}$ with Hamiltonians $H_{s_1}=(-\omega_1/2)\sigma_z$ and $H_{s_2}=(-\omega_2/2)\sigma_z$, where $\omega_1$ and $\omega_2$ denote the self-energies of the qubits $s_1$ and $s_2$, respectively. Finally, the first environment qubit $e_1$, which already played its role in the simulation, is traced out and the next iteration of the evolution continues starting with the upcoming environment qubit $e_2$. It is important to note that there are no intra-environment collisions in this setting. In the present case, the density operator of the bipartite system of two qubits $s_1s_2$ after the $i$th iteration step is given by
\begin{equation}
\rho_{s_1s_2}[i+1]=\Tr_{e_i }\left\{ U_{s_2}  U_{s_1}  U_c(\gamma_{s_2e})  U_c(\gamma_{s_1e}) \rho_{s_1s_2e}[i] U^\dagger_c(\gamma_{s_1e})U^\dagger_c(\gamma_{s_2e}) U^\dagger_{s_1}U^\dagger_{s_2} \right\},
\end{equation}
where $\rho_{s_1s_2e}[i] =\rho_{s_1s_2}[i] \otimes \rho_{e_{i}}[i]$. Note that, in the above equation, while the first interaction operator $U_c(\gamma_{se})$ acts on the system qubit $s_1$ and the $i$th environment qubit $e_i$, the second one affects the qubits $s_2$ and $e_i$ with identical interaction strength, i.e., $\gamma_{s_1e}=\gamma_{s_2e}=\gamma_{se}$.

Recall that our main aim in this work is to compare the effect of coherent PSWAP with that of the incoherent CSWAP during transient dynamics. With this in mind, we also would like to investigate the cases of both system qubits interact with the environment via a CSWAP gate, which we dub as CSWAP-CSWAP model, and a hybrid model where $s_1-e_i$ interaction is modelled by a CSWAP while $s_2-e_i$ interaction is a PSWAP, i.e., a CSWAP-PSWAP model. There is obviously a fourth possibility in which one can consider the swapped application of CSWAP and PSWAP gate in the latter case, however we do not present it here, as it does not generate any different result from the one we consider.


\begin{figure}[t]
\centering
\includegraphics[width=0.85\columnwidth]{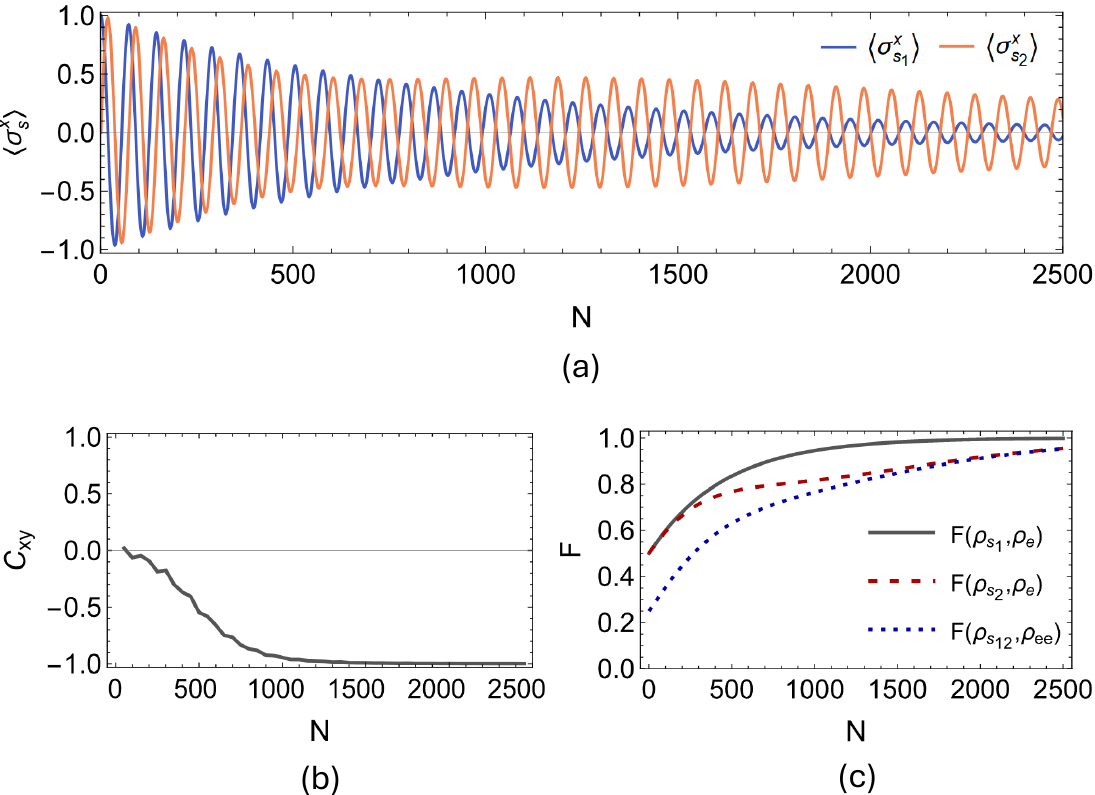}
\caption{System particles are initialized as $\rho_{s_1s_2}=|+\rangle|L\rangle \langle L| \langle+|$ and resonant such that $\omega_1\!=\!\omega_2\!=\!1$, and both interact with a common environmental unit through a coherent PSWAP having strength $\gamma_{se}=0.03(\pi/2)$. While (a) displays the dynamics of $\langle \sigma^x_{s_1} \rangle$ and $\langle \sigma^x_{s_2} \rangle$, (b) shows the corresponding Pearson coefficient $C_{xy}$ between these two data sets settling to $-1$ signaling anti-synchronization, which is plotted considering data windows of 100 collisions with partial overlaps of 50 collisions for N=2500. In (c), we show the fidelity $F$ between the state of the environmental units $\rho_{e}$ and both the local states of the system particles $\rho_{s_1}$, $\rho_{s_2}$ and their global state $\rho_{s_1s_2}$. We observe a clear convergence towards $F=1$, indicating homogenization of system particles with the environment.}
\label{fig5}
\end{figure}

\begin{figure}[t]
\centering
\includegraphics[width=0.85\columnwidth]{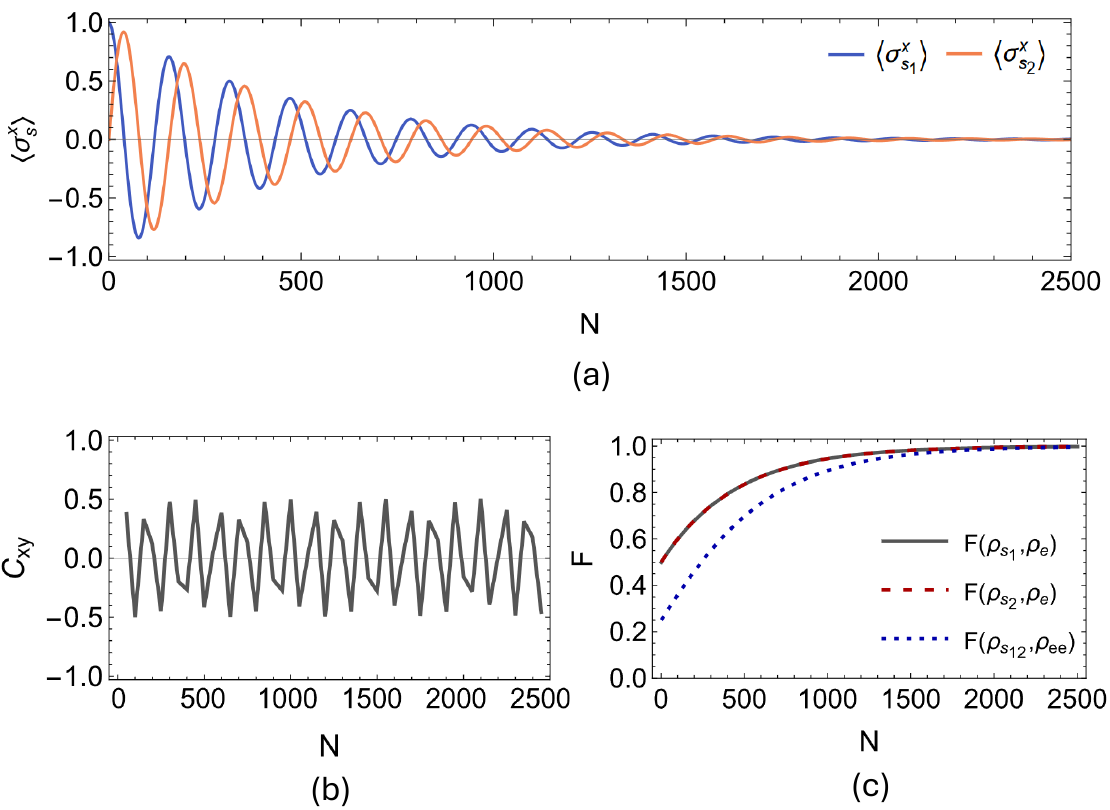}
\centering
\caption{System particles are initialized as $\rho_{s_1s_2}=|+\rangle|L\rangle \langle L| \langle+|$ and resonant such that $\omega_1\!=\!\omega_2\!=\!1$, and both interact with a common environmental unit through a incoherent CSWAP having strength $\gamma_{se}=0.03(\pi/2)$. While (a) displays the dynamics of $\langle \sigma^x_{s_1} \rangle$ and $\langle \sigma^x_{s_2} \rangle$, (b) displays the corresponding Pearson coefficient $C_{xy}$ between these two data sets, which is plotted considering data windows of 100 collisions with partial overlaps of 50 collisions for N=2500, showing no sign of synchronization. In (c), we show the fidelity $F$ between the state of the environmental units $\rho_{e}$ and both the local states of the system particles $\rho_{s_1}$, $\rho_{s_2}$ and their global state $\rho_{s_1s_2}$. We observe a clear convergence towards $F=1$, which indicates homogenization of system particles with the environment.}
\label{fig6}
\end{figure}

Next, let us elaborate on what we mean by spontaneous quantum synchronization and how we will characterize it. The phenomenon of environment-induced spontaneous synchronization is said to emerge between a pair of quantum systems, which are evolving under the effect of an external environment, when coherent phase-locked oscillations in the expectation values of their local observables is established. In order to quantify the degree of synchronous behavior between two qubits, we utilize the widely-used Pearson coefficient, which is typically employed to measure the association between two given data sets. For a pair of discrete variables $x$ and $y$, the Pearson coefficient $C_{xy}$ quantifies the degree of linear correlation between them, which is mathematically expressed as
\begin{equation}
C_{xy}=
\frac{\sum_n(x_n-\bar{x})(y_n-\bar{y})}{\sqrt{\sum_n(x_n-\bar{x})^2}\sqrt{\sum_n(y_n-\bar{y})^2}},
\label{eq:pc}
\end{equation} 
where $\bar{x}$ and $\bar{y}$ are the averages of the two variables $x$ and $y$ over the data set of length $n$. The Pearson coefficient $C_{xy}$ can assume values from the interval $[-1, 1]$. While $C_{xy}=-1$ corresponds to a completely negative correlation between the considered pair of variables, $C_{xy}=1$ means that a completely positive correlation exists between them. In our treatment, the considered variables are the expectation values of local spin observables for the open system qubits $s_1$ and $s_2$, more specifically $\langle \sigma^x_{s_1} \rangle=\Tr{\sigma_x \rho_{s_1}}$ and $\langle \sigma^x_{s_2}\rangle=\Tr{\sigma_x \rho_{s_2}}$, where $\rho_{s_1}=\Tr_{s_2}\left\{\rho_{s_1 s_2}\right\}$ and $\rho_{s_2}=\Tr_{s_1}\left\{\rho_{s_1 s_2}\right\}$. Therefore, form the viewpoint of synchronization, completely positive (negative) correlation implies that there emerges a fully synchronized (anti-synchronized) behavior between the local expectation values of the system qubits. Indeed, in general it is said that phase-locking between the oscillations in the expectation values of the spins is achieved when the Pearson coefficient stabilizes at a constant value, a phenomenon known as time-delayed synchronization. At this point, we should also emphasize that even though we focus on a particular spin observable for concreteness, i.e. $\sigma_x$, the absence or existence of spontaneous mutual synchronization in our analysis is indeed independent of the chosen observable. Rather, it is a dynamical property of the considered model. In our simulations, we calculate the expectation values $\langle \sigma^x_{s_1}\rangle$  and $\langle \sigma^x_{s_2}\rangle$, and evaluate the Pearson coefficient $C_{xy}$ over a sliding data window along the total collision number $N$, which enables us to trace the time-evolution of the Pearson coefficient and thus look for any phase-locked oscillations during the dynamics. Additionally, we allow partial overlapping of adjacent data windows for some fixed data interval to present smoother plots of Pearson coefficient in terms of the collision number.

Here, for all the considered two-qubit collision models with different interactions schemes, we set the initial state of the open qubit pair as $\rho_{s_1s_2}=\rho_{s_1}\otimes \rho_{s_2}=|+\rangle|L\rangle \langle L| \langle+|$, where $|L\rangle=(1/\sqrt{2})(|0\rangle + i|1\rangle)$. Furthermore, the system-environment interaction strength is $\gamma_{se}=0.03(\pi/2)$ for both $s_1$ and $s_2$ and the self-energies of the system qubits are considered to be equal, $\omega_1=\omega_2=1$, meaning that they are resonant with $\delta t=0.04$. We should stress that we choose these initial system and the interaction parameters solely for presentation purposes. The phenomenon of environment-induced spontaneous synchronization between the qubit pair $s_1$ and $s_2$ dynamically emerges independently of their initial state. 


\begin{figure}[t]
\centering
\includegraphics[width=0.85\columnwidth]{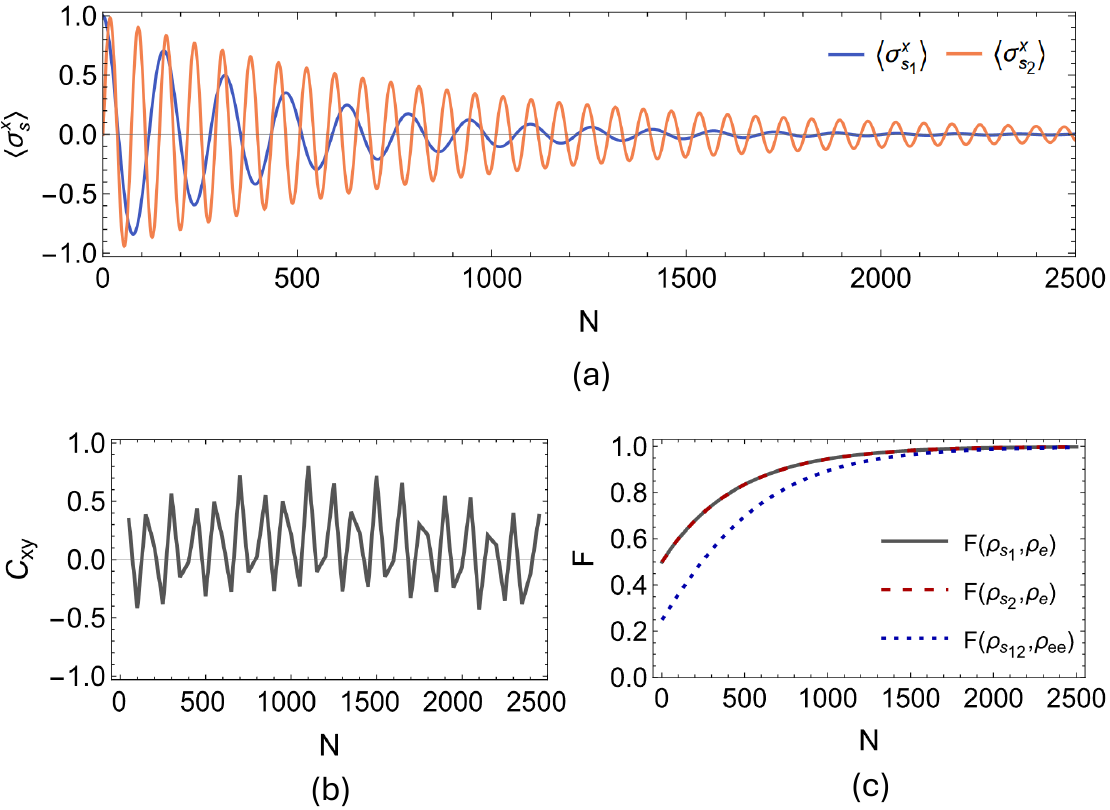}
\caption{System particles are initialized as $\rho_{s_1s_2}=|+\rangle|L\rangle \langle L| \langle+|$ and resonant such that $\omega_1\!=\!\omega_2\!=\!1$. As the first particle interact with environmental units through incoherent CSWAP, the second one interacts with the same environment units via a coherent PSWAP interactions, with identical strength $\gamma_{se}=0.03(\pi/2)$. While (a) displays the dynamics of $\langle \sigma^x_{s_1} \rangle$ and $\langle \sigma^x_{s_2} \rangle$, (b) displays the corresponding Pearson coefficient $C_{xy}$ between these two data sets, which is plotted considering data windows of 100 collisions with partial overlaps of 50 collisions for N=2500, showing no sign of synchronization. In (c), we show the fidelity $F$ between the state of the environmental units $\rho_{e}$ and both the local states of the system particles $\rho_{s_1}$, $\rho_{s_2}$ and their global state $\rho_{s_1s_2}$. We observe a clear convergence towards $F=1$, confirming the homogenization of system particles with the environment.}
\label{fig7}
\end{figure}

First, we present our findings regarding the onset of spontaneous synchronization between the open qubit pair for the two-qubit PSWAP-PSWAP model. In Figure \ref{fig5} (a) and (b), we display the evolution of the expectation values of local spin observables $\langle \sigma^x_{s_1}\rangle$ and $\langle \sigma^x_{s_2}\rangle$ for the PSWAP-PSWAP model for $n=2500$ collisions, and the dynamics of the Pearson coefficient $C_{xy}$, which is calculated by considering data windows of $100$ collisions with partial overlaps of $50$ collisions. It can clearly be observed that the environment-induced synchronization is dynamically established between the open system qubits. In fact, what is established here in case of the PSWAP-PSWAP collision model is complete anti-synchronization between the open system qubits $s_1$ and $s_2$ since the Pearson coefficient eventually stabilizes at $C_{xy}=-1$. In Figures~\ref{fig6} and~\ref{fig7}, we present the outcomes of the same analysis performed for the remaining two interaction settings, that is, for the CSWAP-CSWAP and CSWAP-PSWAP models, respectively. As can be straightforwardly seen from parts (a) and (b) of the plots in Figure~\ref{fig6} and~\ref{fig7}, synchronization cannot be established in either of these two models since no phase-locking occurs between the expectation values of the open system qubits. This observation implies that synchronization is also suppressed by the CSWAP gate. In panel (c) of all Figures~\ref{fig5}-\ref{fig7}, we show the fidelity between the state of the environmental unit, and local and global system states. We clearly observe homogenization of system particles with the environment across all of these cases. Therefore, similar to the previous section on non-Markovianity, even though the dynamical processes take the system from the same initial state to the same final state, the transient time dynamical behavior displays a stark difference between coherent and incoherent homogenizers. That is, the latter suppresses the establishment of transient synchronization between the system qubits, whenever it is it is involved in the description of the dynamics.

A common and natural extension of the synchronization discussion that we presented above is the following: What happens in case the two system qubits have some detuning between their self-energies, i.e. $\omega_1\neq\omega_2$? In fact, the answer to such a question in the quantum synchronization literature is to introduce an interaction between the system particles, which eventually compensates for the breakdown of synchronization for detuned qubits and restores the behavior observed in the resonant case~\cite{Galve2017,Giorgi2019,Karpat2019,Karpat2020,Karpat2021}. Frankly, in the present models it is only possible to restore synchronization with $s_1-s_2$ interactions in case of detuning, if one considers only coherent PSWAP interactions. However, in such a case, we do not observe homogenization of neither global nor local system states with the environment state, regardless of the combinations and order of interactions. In order to be able to make a fair comparative analysis, we do not present these results here. Nevertheless, we believe the mechanisms hindering homogenization for bipartite or multipartite systems may be an interesting direction to investigate.

\section{Conclusion}\label{conc}

In summary, we have presented a comparative analysis of coherent and incoherent collision models, where pairwise couplings between the model constituents are respectively realized by PSWAP and CSWAP  operators, which are universal homogenizers in the Markovian limit, from several different perspectives. We began our discussion considering the Markovian dynamics of a single qubit  interacting with a stream of environment qubits, which do not undergo intra-environment interactions. In this case, we showed that although the asymptotic behavior of the open system qubit under both PSWAP and CSWAP based collision models are the same, its transient behavior turned out to be significantly dissimilar as we demonstrated by visualizing its evolution path on the Bloch sphere.

In order to investigate the emergence of non-Markovian memory effects in the dynamics, we then focused on single qubit collision models also having interactions between consecutive environment qubits, in addition to the system-environment coupling. In this scenario, we revealed that memory effects in the dynamics of the open system qubit emerge only if the interaction between the open system and environment qubits is coherent, that is, described by the PSWAP operator. We also showed that, when the system-environment coupling is coherent, incoherent intra-environment interactions realized by the CSWAP operator can result in greater degree of non-Markovianity as compared to the case of having coherent PSWAP intra-environment interactions, assuming identical coupling strengths. Next, we turned our attention to two-qubit coherent and incoherent collision models to study the onset of spontaneous synchronization between the qubits. In this setting, there only exist system-environment interactions, i.e., intra-environment couplings are not taken into account. We established that phase-locking and thus synchronous behavior between the expectation values of the local spin observables could only be observed when the system-environment interactions are realized by the coherent PSWAP operator.

We believe that our work offers a detailed look into homogenization of quantum systems through coherent PSWAP and incoherent CSWAP interactions. Even though the process of quantum homogenization may be viewed as a straightforward way of relaxing the system state to that of the environment by sequential interactions, we show that the actual dynamical path taken during this process drastically change the characteristics of the process that develop in the transient time. Additionally, we only know that Markovian PSWAP, CSWAP, and non-Markovian PSWAP-PSWAP interactions are universal homogenizers~\cite{Ziman2002,Beever2024,Saha2024}. An interesting direction is to extend these results to different non-Markovian scenarios and, especially, to multipartite system settings. We anticipate that classical and/or quantum correlations among the building blocks of the collision model will play an important role~\cite{Comar2021,Filippov2021}. Both the analysis of differences in transient time dynamics and universality of quantum homogenizers can also have different implications in relation to recently introduced homogenization based quantum state preparation proposals~\cite{Yosifov2024,Yosifov2024arXiv}.


\vspace{20pt} 

\authorcontributions{All authors contributed equally to this work. All authors have read and agreed to the published version of the~manuscript.}

\funding{G.K. is supported by The Scientific and Technological Research Council of T\"{u}rkiye (TUBITAK) through the 100th Anniversary Incentive Award.}


\conflictsofinterest{The authors declare no conflict of interest.} 


\reftitle{References}

\bibliography{biblio}

\end{document}